\documentclass[reprint,superscriptaddress,showpacs,amsmath,amssymb,prb]{revtex4-2}
\usepackage{natbib}
\usepackage{siunitx}
\usepackage{array}
\usepackage{xcolor}
\usepackage{subfigure} 
\usepackage{amsmath} 
\usepackage{amsfonts}
\usepackage{amssymb}
\usepackage{graphicx}
\usepackage{hyperref}
\usepackage{appendix}
\usepackage{academicons}
\usepackage{orcidlink}

\providecommand{\vect}[1]{{\boldsymbol{#1}}}

\usepackage{physics}

\begin{document} 
\title{Spin-Seebeck Signatures of Spin Chirality in Kagome Antiferromagnets}

\author{Feodor Svetlanov Konomaev}
\affiliation{Department of Engineering Sciences, University of Agder, 4879 Grimstad, Norway} 
\author{Mithuss Tharmalingam}
\affiliation{Department of Engineering Sciences, University of Agder, 4879 Grimstad, Norway} 
\author{Kjetil M. D. Hals}
\affiliation{Department of Engineering Sciences, University of Agder, 4879 Grimstad, Norway} 
\date{\today}
\newcommand{\Kjetil}[1]{\textcolor{red}{#1}} 
\begin{abstract}
Non-collinear antiferromagnets (NCAFMs) are appealing for antiferromagnetic spintronics, as they combine the advantages of collinear antiferromagnets with novel emergent phenomena stemming from their complex spin structures. These phenomena are often associated with the intrinsic spin chirality, which characterizes the handedness of the ground-state spin configuration. Here, we investigate a kagome NCAFM interfaced with a normal metal and demonstrate that the ground-state vector spin chirality can be probed through measurements of the spin Seebeck effect (SSE). Starting from a microscopic spin Hamiltonian, we derive the corresponding bosonic Bogoliubov-de Gennes Hamiltonians for the two chiral configurations. Using linear response theory, we obtain a general expression for the spin current thermally pumped into the normal metal by the SSE. We show that a sizable in-plane spin current emerges exclusively in the negative-chiral state, providing a direct signature for real-time detection of chirality switching in kagome NCAFMs. In addition, we find a field-dependent out-of-plane spin current whose magnitude differs between the two chiralities by about 81\%, reflecting their distinct magnon band structures.
\end{abstract}

\maketitle 

\section{Introduction} 
The intrinsic spin chirality of a magnetic state reflects the handedness of spin arrangements in non-collinear or non-coplanar systems. This handedness is commonly characterized by a vector or scalar chirality, which captures the relative orientation of neighboring spins on the lattice. Crucially, spin chirality has been shown to play a central role in the emergence of topological phases and in driving unconventional transport phenomena~\cite{Sitte:jap2018,Bobel:pr2021,Manchon:ssp2017, Brataas:nn2014}.

\begin{figure}[h] 
\centering 
\includegraphics[trim=10 0 3.5 130,clip,scale=1.1]{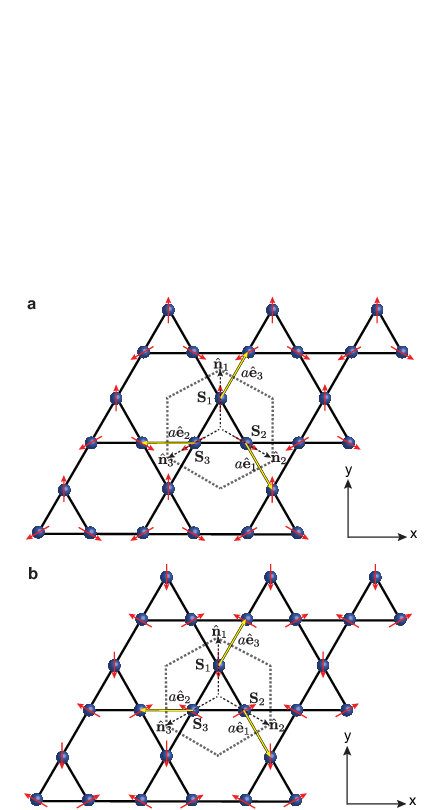}  
\caption{(color online). A schematic representation of a kagome NCAFM monolayer in a ground state with {\bf a}. positive and {\bf b}. negative chirality. The easy axes are denoted with dashed black arrows, the spin directions with red arrows, and the boundaries of the unit cell are outlined with dashed grey lines, respectively.}
\label{Fig1} 
\end{figure} 

Chiral magnetic orders have garnered significant attention in the study of non-collinear antiferromagnets (NCAFMs) --- antiferromagnetic systems in which the spin sublattices are ordered non-collinearly \cite{Andreev:spu1980}. The complexity and chirality of their magnetic states are believed to give rise to unconventional magnetic, electronic, and optical phenomena, even in the absence of a net magnetization~\cite{Smejkal:nrm2022, Nakatsuji:nature2015, Nayak:Sadv2016, Kiyohara:pra2015, Ikhlas:nphys2017, Higo:np2018, Kimata:n2019, Sallermann:prl2022}. A particularly attractive subclass of NCAFMs for exploring chirality-driven spin phenomena is the kagome antiferromagnets (AFMs). Kagome AFMs are two-dimensional (2D) spin systems composed of corner-sharing triangles and are considered among the most geometrically frustrated 2D spin lattices~\cite{Grohol:nm2005} (see Fig.~\ref{Fig1}). Important examples of materials that can be characterized as kagome AFMs are ultrathin films of the Weyl antiferromagnets~\cite{Kuroda:natmat2017} $\text{Mn}_3\text{X}$ (X = Ge, Sn, Ga, Ir) and iron jarosites~\cite{Grohol:nm2005}, which consist of stacked kagome layers. In equilibrium, the three sublattice spins $\vect{S}_1$, $\vect{S}_2$, and $\vect{S}_3$ on each triangle are oriented at 120° to one another. The handedness of this spin configuration is described by the vector spin chirality $\vect{K}_\mathrm{v}=\vect{S}_2\times\vect{S}_1+\vect{S}_3\times\vect{S}_2+\vect{S}_1\times\vect{S}_3$~\cite{Grohol:nm2005}. For coplanar arrangements, $\vect{K}_\mathrm{v}$ points parallel or antiparallel to the z-axis, corresponding to the ground states of $(+)$-chirality and $(-)$-chirality, respectively (see Fig.~\ref{Fig1}). When subjected to an external magnetic field or an in-plane Dzyaloshinskii-Moriya interaction (DMI), the sublattice spins acquire a slight out-of-plane canting, leading to a non-coplanar spin texture. In this regime, the spin configuration also exhibits a finite scalar spin chirality expressed through the quantity $K_\mathrm{s}=\vect{S}_1\cdot(\vect{S}_3\times\vect{S}_2)$~\cite{Grohol:nm2005}.

Kagome AFMs display unique properties that make them highly promising for applications in antiferromagnetic spintronics~\cite{Jungwirth:np2018, Duine:np2018, Gomonay:np2018, Zelezny:np2018, Nemec:np2018, Libor:np2018, Baltz:rmp2018}.  
Notably, experiments have demonstrated large spin Hall~\cite{Kimata:n2019}, anomalous Hall~\cite{Nakatsuji:nature2015, Nayak:Sadv2016, Kiyohara:pra2015}, and anomalous Nernst effects~\cite{Ikhlas:nphys2017} at room temperature, which are attributed to the finite Berry curvature arising from the chiral spin texture. Additionally, recent theoretical and experimental studies have demonstrated that the  spin order in these systems can be controlled through current-induced torques~\cite{Gomonay:PRB2012, Tserkovnyak:PRB2017, Ochoa:PRB2018, Yamane:PRB2019, Li:PRB2021, Lund:PRB2021, Rodrigues:PRB2022, Lund:PRB2023, Lee:nn2025, Takeuchi:nmat2021, Deng:nsr2023, Tsai:nature2020, Higo:Nature2022,Yoon:nmat2023,Xie:nc2022,Yan:am2022}, strain~\cite{Tharmalingam:prb2025}, and laser pulses~\cite{Reichlova:nc2019}. In textured kagome AFMs, theoretical studies have predicted unconventional coupling mechanisms between spin waves and domain walls~\cite{Rodrigues:PRL2021}, as well as the emergence of p-wave magnetism~\cite{Chakraborty:nc2025}. Moreover, kagome AFMs have been shown to host topological magnons~\cite{Laurell:PRB2018, Hals:PRB2025}, which manifest as chiral edge modes.

Another intriguing feature of kagome AFMs is the strong dependence of their magnon excitation spectrum on the vector chirality of the ground state. This was recently demonstrated in studies of auto-oscillations, where it was shown that the $(-)$-chiral ground state supports gapless self-oscillations, while the $(+)$-chiral state exhibits a finite gap \cite{Lund:PRB2021}. However, the broader implications of this vector chirality dependence for other observable phenomena remain largely unexplored. In this work, we show that the spin Seebeck effect (SSE)~\cite{Bauer:review}
 -- the generation of spin currents by temperature gradients -- can serve as an experimental probe of the vector chirality of kagome AFMs. When interfaced with a heavy metal, magnons in the kagome AFM thermally pump a spin current into the adjacent metal under a temperature gradient across the interface. Our results reveal that both the polarization and magnitude of the spin current are strongly influenced by the vector chirality of the magnetic state. This finding establishes the SSE as a promising mechanism for detecting chirality switching in chiral antiferromagnets.

This paper is organized as follows. In Sec.~\ref{Sec2}, we present the microscopic model of the kagome AFM and determine the magnon band structure for the two ground states with opposite vector chirality. Section~\ref{Sec3} is devoted to the calculation of the SSE for each chirality, along with an analysis of the resulting differences in their SSE responses. In Sec.~\ref{Sec4}, we discuss possible experimental signatures of the vector chirality in the inverse spin Hall voltage. Concluding remarks are provided in Sec.~\ref{Sec5}.

\section{Theory}\label{Sec2}
The kagome AFM is described by the Hamiltonian~\cite{Rodrigues:PRB2022} 
\begin{equation}
\mathcal{H}_\text{AF}=\mathcal{H}_\text{E}+\mathcal{H}_\mathrm{A}+\mathcal{H}_\mathrm{DM}+\mathcal{H}_\mathrm{ext}\text{,}
 \label{eq:ham}
\end{equation}
which includes the following contributions: an exchange interaction between nearest-neighbor spins $\vect{S}_i$ and $\vect{S}_j$ with coupling strength $J>0$, represented by $\mathcal{H}_\text{E}=J\sum_{\langle ij\rangle} \vect{S}_i\cdot\vect{S}_j$, 
easy-plane ($K_\perp>0$) and easy-axis ($K>0$) anisotropy energies, characterized by $\mathcal{H}_\mathrm{A}=\sum_i \left(K_\perp(\vect{S}_i\cdot\hat{\vect{z}})^2-K(\vect{S}_i\cdot\hat{\vect{n}}_i)^2\right)$, 
an out-of-plane DMI between nearest neighbors in the form of $\mathcal{H}_\mathrm{DM}=\sum_{\langle i, j\rangle}D \vect{\hat{z}}\cdot(\vect{S}_i\times\vect{S}_j)$ with $\langle i,j\rangle \in \{ \langle 1,3\rangle, \langle 2,1\rangle, \langle 3,2\rangle \}$~\cite{Rodrigues:PRB2022}, 
and the Zeeman coupling to an external magnetic field $B$ applied perpendicular to the kagome lattice plane, $\mathcal{H}_\mathrm{ext}=-\sum_i\gamma B\hat{\vect{z}}\cdot\vect{S}_i$, where $\gamma$ is the gyromagnetic ratio. 
The in-plane easy-axes at the three spin sub-lattice sites are $\hat{\vect{n}}_1=[0,1,0]^T$, $\hat{\vect{n}}_2=[\frac{\sqrt{3}}{2},-\frac{1}{2},0]^T$, and $\hat{\vect{n}}_3=[-\frac{\sqrt{3}}{2},-\frac{1}{2},0 ]^T$, respectively.
In what follows, it will often be convenient to represent a lattice site as $i \equiv (\kappa, \alpha)$, with $\kappa$ denoting the unit-cell index and $\alpha \in\{1,2,3\}$ the sublattice index.

The vector spin chirality of the ground state is governed by the ratio between the out-of-plane DMI and the in-plane easy-axis anisotropy. 
For $D/K < 1/4\sqrt{3}$ and $B=0$, the three sub-lattice spins $\vect{S}_1$, $\vect{S}_2$, and $\vect{S}_3$ within the unit cell align parallel or antiparallel to the unit vectors $\tilde{\vect{n}}_1^+=\hat{\vect{n}}_1$, $\tilde{\vect{n}}_2^+=\hat{\vect{n}}_2$ and $\tilde{\vect{n}}_3^+=\hat{\vect{n}}_3$, respectively. 
We refer to this configuration as the $(+)$-chiral state, characterized by a vector spin chirality $\vect{K}_\mathrm{v}\propto + \hat{\vect{z}}$. 
For $D/K > 1/4\sqrt{3}$ and $B=0$, the sub-lattice spins instead align parallel or antiparallel to $\tilde{\vect{n}}_1^-=(0,-1,0)$, $\tilde{\vect{n}}_2^-=(\cos(\pi/6+\alpha),\sin(\pi/6+\alpha),0)$ and $\tilde{\vect{n}}_3^-=(-\cos(\pi/6+\alpha),\sin(\pi/6+\alpha),0)$. 
In this case, the vector spin chirality is $\vect{K}_\mathrm{v}\propto -\hat{\vect{z}}$ and the ground state is referred to as the $(-)$-chiral state.
Owing to the in-plane easy-axis anisotropy, the vectors $\tilde{\vect{n}}_{2,3}^-$ acquire a small tilt toward the $y$-direction by an angle $\alpha$. This deviation perturbs the ideal $120^\circ$ sublattice spin configuration, giving rise to a weak net in-plane spin polarization.
An explicit expression for $\alpha$ is given in Appendix~\ref{app:alpha}.
In the presence of a magnetic field, the spins additionally acquire a small out-of-plane tilting by an angle $\theta$, the explicit expression for which is provided in Appendix~\ref{app:theta}.
This tilting leaves the handedness of the spins unchanged but generates a net out-of-plane spin polarization of the thermally excited magnons, which is essential for a finite SSE in the $(+)$-chiral state, as we show in Sec.~\ref{Sec3}.

The collective spin excitations of the above Hamiltonian can be obtained via the Holstein-Primakoff transformation~\cite{Auerbach:book}, which represents the spin operators in terms of bosonic ladder operators $a_i$ and $a_i^\dagger$:
$S_{i,\tilde{x}_i^\pm}=\hbar\sqrt{\frac{S}{2}}(a_i+a_i^\dagger)$, $S_{i,\tilde{y}_i^\pm}=\frac{\hbar}{i}\sqrt{\frac{S}{2}}(a_i-a_i^\dagger)$, and $S_{i,\tilde{z}_i^\pm}=\hbar(S-a_i^\dagger a_i)$.
Here, $\tilde{\vect{x}}_i^\pm= \tilde{\vect{n}}_i^\pm\cross\hat{\vect{z}}$, $\tilde{\vect{y}}_i^\pm= \tilde{\vect{z}}_i^\pm\cross\tilde{\vect{x}}_i^\pm$, and $\tilde{\vect{z}}_i^\pm=\vect{S}_i^{(0)\pm}/\hbar S$ define the local reference frame at lattice site $i$ such that the quantization axis $\tilde{\vect{z}}_i^\pm$ points along $\vect{S}_i^{(0)\pm}$. The expressions for the spin operators are substituted into the Hamiltonian \eqref{eq:ham} and transformed to momentum space using the Fourier transformation $a_i = (1/\sqrt{N_{\rm AF}})\sum_{\vect{q}} a_{\alpha, \vect{q}} \exp (i\vect{R}_{\kappa, \alpha}\cdot\vect{q})$, where $N_{\rm AF}$ is the number of magnetic unit cells, and $\vect{R}_{\kappa, \alpha}\equiv \vect{R}_\kappa + \vect{\delta}_\alpha$ denotes the position of spin $i$, expressed in terms of the unit cell position vector $\vect{R}_\kappa$ and the vector $\vect{\delta}_\alpha$ specifying the position of sub-lattice spin $\alpha$ in the unit cell. The above substitution and transformation map the spin Hamiltonian~\eqref{Fig1}  onto a bosonic Bogoliubov-de-Gennes (BdG) Hamiltonian for the ladder operators
\begin{equation}\label{eq:bdg}
    \mathcal{H}_\text{AF}=\frac{1}{2}\sum_{\vect{q}} (\vect{\alpha}_{\vect{q}}^{\dagger}, \vect{\alpha}_{-\vect{q}} )\mathbf{H}_{\vect{q}} \begin{pmatrix} \vect{\alpha}_{\vect{q}}\\ \vect{\alpha}_{-\vect{q}}^{\dagger} \end{pmatrix} \text{,}
    \end{equation}
where $\vect{\alpha}_{\vect{q}}^\dagger=(a_{1,\vect{q}}^\dagger,a_{2,\vect{q}}^\dagger,a_{3,\vect{q}}^\dagger)$.
Throughout, $\vect{q}$ denotes wave vectors in the first Brillouin zone (1BZ). The full $6\times 6$ matrix form of $\mathbf{H}_{\vect{q}}$ is given in Appendix~\ref{app:bdg}. 

We diagonalize the BdG Hamiltonian~\eqref{eq:bdg} numerically by a  
$6\times6$ paraunitary matrix $\mathbf{T}_{\vect{q}}^{-1}= [ \mathbf{U}_{\vect{q}} , \mathbf{V}_{-\vect{q}}^* ;   \mathbf{V}_{\vect{q}} , \mathbf{U}_{-\vect{q}}^*] $, where $\mathbf{U}_{\vect{k}} $ and $\mathbf{V}_{\vect{q}} $ are $3 \times3$ submatrices~\cite{Colpa}: 
\begin{equation}\label{eq:transform}
    (\mathbf{T}_{\vect{q}}^{-1})^\dagger \mathbf{H}_{\vect{q}}\mathbf{T}_{\vect{q}}^{-1}=\mathrm{diag}(\mathbf{E}_{\vect{q}},\mathbf{E}_{-\vect{q}})\text{.}
    \end{equation}
In the above expression, $\mathbf{E}_{\vect{q}}=\mathrm{diag}\left(\varepsilon_3(\vect{q}),\varepsilon_2(\vect{q}),\varepsilon_1(\vect{q})\right)$ with $\varepsilon_{n}(\vect{q})$ denoting the eigenenergy of the $n$-th band.
The diagonalized form of Eq.~\eqref{eq:bdg} is $\mathcal{H}=\sum_{n,\vect{q}} \varepsilon_n (\vect{q}) \gamma^{\dagger}_{n,\vect{q}} \gamma_{n,\vect{q} }$, where  $\gamma^{\dagger}_{n,\vect{q}}$ and $\gamma_{n,\vect{q} }$
are the ladder operators describing the elementary magnonic spin excitations of the AFM.

\begin{figure}[ht] 
\centering 
\includegraphics[trim=80 105 100 100,clip,scale=0.57]{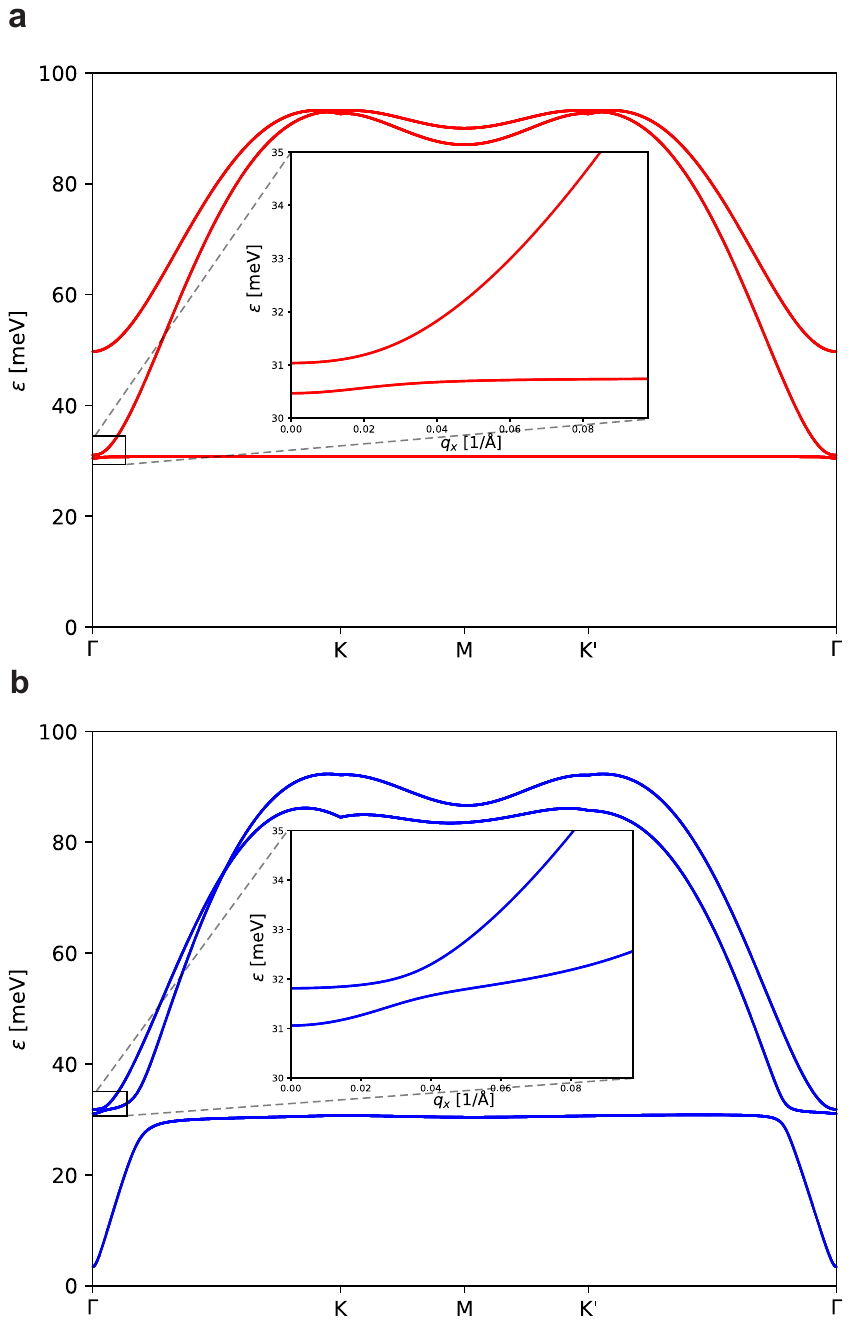}  
\caption{(color online). Magnon energy bands along high-symmetry lines in the first Brillouin zone for {\bf a}. the $(+)$-chiral ground state and {\bf b}. the $(-)$-chiral ground state. The inset diagrams highlight the band structures near the $\Gamma$ point along the $x$-axis.
We have used the  parameter values $\sqrt{2}a=3.785$~Å, spin $S=1$, $\gamma=1.76\cdot 10^{11}$~1/Ts, $J=39.94~\mathrm{meV}/\hbar^2$, $K=5~\mathrm{meV}/\hbar^2$, $K_{\bot}=0$, $D = K/4\sqrt{3}$, and $B = 5$~T. }
\label{Fig2} 
\end{figure} 

Fig.~\ref{Fig2} shows the magnon band structures calculated from Eq.~\eqref{eq:transform} for the ($\pm$)-chiral states. The parameters correspond to $\text{Mn}_3\text{Ir}$ with lattice spacing $\sqrt{2}a=3.785$~Å \cite{Szunyogh:PRB2009}, spin $S=1$, gyromagnetic ratio $\gamma=1.76\cdot 10^{11}$~1/Ts, $J=39.94~\mathrm{meV}/\hbar^2$~\cite{Jenkins:JAP2018, Szunyogh:PRB2009}, $K_{\bot}=0$. $\text{Mn}_3\text{Ir}$ possesses three effective easy axes oriented approximately along the directions shown in Fig.~\ref{Fig1} \cite{Szunyogh:PRB2009} with an associated effective anisotropy parameter of $K\approx 5~\mathrm{meV}/\hbar^2$~\cite{Jenkins:JAP2018, Szunyogh:PRB2009}.  For a $\text{Mn}_3\text{Ir}$ monolayer oriented along the kagome plane, these easy axes are expected to lie within the plane and coincide with those assumed in Eq.~\eqref{eq:ham}.
For the out-of-plane DMI, we take the threshold value corresponding to the transition between positive and negative chiralities, $D = K/4\sqrt{3}$. 
Unless otherwise specified, these material parameters are used in all subsequent calculations. 
Additionally, for the band structures in Fig.~\ref{Fig2}, an external magnetic field of $B = 5$~T is applied.

The main difference between the band structures of the $(\pm)$-chiral states arises near the $\Gamma$ point, as highlighted in the inset diagrams of Fig.~\ref{Fig2}. In the $(+)$-chiral state, the lowest band exhibits a finite energy gap at $\vect{q}=0$, whereas in the $(-)$-chiral state the lowest band approaches zero as $\vect{q}\to 0$. At $\vect{q}=0$, the band retains only a small finite value determined by the in-plane tilt angle $\alpha$, which vanishes in the strong exchange-coupling limit considered in Ref.~\cite{Lund:PRB2023, Comment}. 
At finite temperature, this results in a larger magnon population in the lowest band of the $(-)$-chiral state compared to the $(+)$-chiral state, thereby leading to an amplification of the SSE, as we demonstrate below. 

\section{The Spin Seebeck Effect}\label{Sec3}
\begin{figure}[ht] 
\centering 
\includegraphics[scale=0.7]{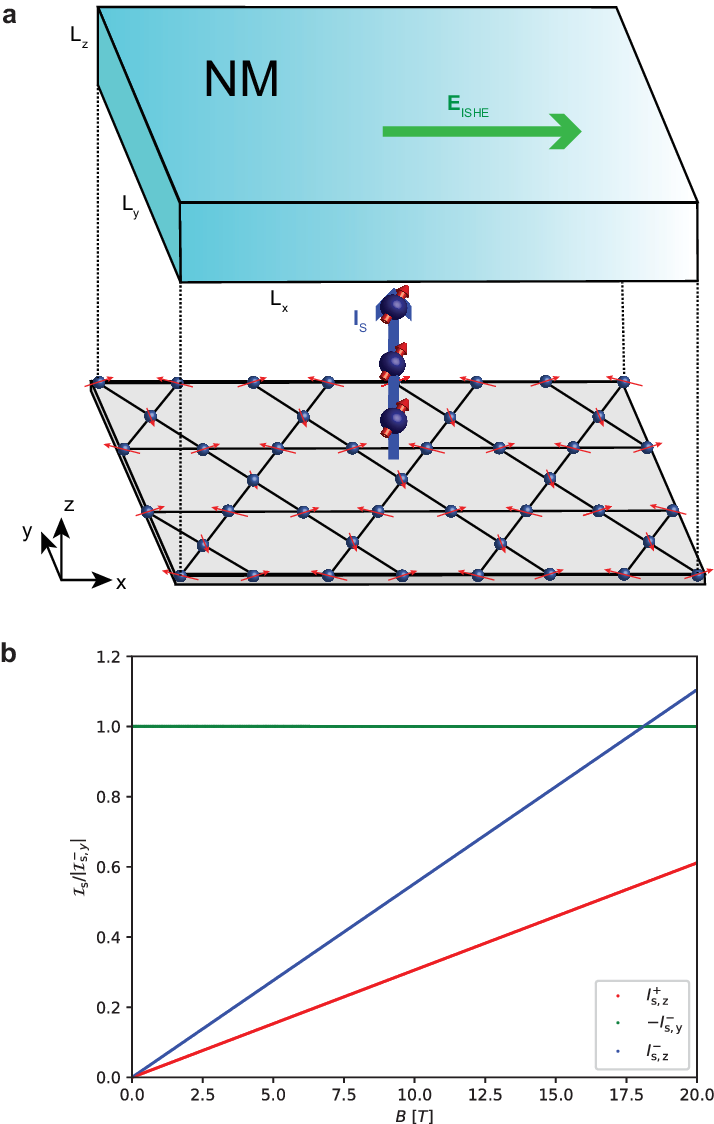}  
\caption{(color online). {\bf a}. Schematic of a 2D kagome AFM/3D NM heterostructure (NM in cyan). The SSE-driven spin current (blue arrow) yields an ISHE-induced electric field (green arrow) in the NM.
{\bf b}. The $z$-component of the spin current in the positive (red) and negative (blue) chiral states, normalized by the constant $y$-component of the spin current in the negative chiral state (green), as a function of the applied out-of-plane magnetic field $B$.  
In both cases, $I_{\mathrm{s},x}=0$. 
Results are shown for $T_\mathrm{AF}=300$~K, $T_\mathrm{N}=299$~K, $\tau_s=50$~fs, $\lambda=5.21$~nm, and the AFM parameters as specified in Sec.~\ref{Sec2}. }
\label{Fig3} 
\end{figure}

Next, we examine the influence of the ground-state chirality on the SSE when a 2D kagome AFM is interfaced with a three-dimensional (3D) normal metal (NM), see Fig.~\ref{Fig3}a.
The Hamiltonian of the heterostructure is
\begin{equation}\label{eq:totall}
\mathcal{H}=\mathcal{H}_\mathrm{NM} + \mathcal{H}_\mathrm{AF}+ \mathcal{H}_I \text{.}
\end{equation}
The isolated AFM is described by the Hamiltonian in Eq.~\eqref{eq:bdg}, while the itinerant charge carriers in the NM are governed by the Hamiltonian 
$\mathcal{H}_\mathrm{NM}=\sum_{\vect{k},\tau} \epsilon_{\vect{k}} c_{\vect{k}\tau}^\dagger c_{\vect{k}\tau}$,  where $c_{\vect{k}\tau}^\dagger$ creates a particle with momentum $\hbar \vect{k}$ and spin $\tau$, and $\epsilon_{\vect{k}}$ denotes its energy.
The interfacial exchange interaction between the carrier spin density $\vect{s}(\vect{r})= \frac{\hbar}{2} \Psi_\tau^\dagger(\vect{r})\boldsymbol{\sigma}_{\tau\tau^{'}}\Psi_{\tau^{'}} (\vect{r})$
and the AFM spins $\vect{S}_{i}$ at the AFM/NM interface ($I$) is
\begin{equation}\label{eq:interaction}
\mathcal{H}_I= \sum_{i\in I} \int d^3r J_i \rho_{i} (\vect{r})\vect{S}_{i}\cdot\vect{s}(\vect{r})\text{,}
 \end{equation}
where $\rho_{i}(\vect{r})$ is the probability density of the localized AFM spin $\vect{S}_{i}$, $J_i$ its exchange coupling to the carriers, and
$\boldsymbol{\sigma}$ is a vector consisting of the Pauli matrices. 

The fermionic field operator describing the NM can be expanded in a Wannier basis as $\Psi_\tau^{\dagger}(\vect{r})=\sum_{n\in NM}\Psi_{\vect{R}_n}^{\ast}(\vect{r})c_{n\tau}^{\dagger}$, 
where $c_{n\tau}^{\dagger}$ creates an electron with spin $\tau$ at the lattice site $\vect{R}_n$ in the NM, and $\Psi_{\vect{R}_n}^{\ast}(\vect{r})$ denotes the Wannier function localized at $\vect{R}_n$.  
Focusing on low-energy excitations near the Fermi level, we restrict to a single band. Substituting the expansion into Eq.~\eqref{eq:interaction} and assuming that each AFM spin $\vect{S}_{i}$ overlaps with a single Wannier orbital $\Psi_{\vect{R}_{i}}(\vect{r})$, we obtain the tight-binding form of the interfacial Hamiltonian,
\begin{equation}\label{eq:discreteinteraction}
\mathcal{H}_I=\sum_{i\in I}J_{\mathrm{sd}, i} \vect{S}_{i}\cdot\vect{s}_{i},
\end{equation}
with $J_{\mathrm{sd},i} \equiv \int {\rm d}\vect{r} J_i \rho_i(\vect{r}) |\Psi_{\vect{R}_{i}}(\vect{r})|^2$ and $\vect{s}_i=(\hbar/2)c_{i,\tau}^\dagger\boldsymbol{\sigma}_{\tau\tau'}c_{i,\tau'}$. The lattice operators $c_{n\tau}$ relate to $c_{\vect{k}\tau}$ in $\mathcal{H}_\mathrm{NM}$ via the Fourier transform $c_{n\tau}=(1/\sqrt{N_N})\sum_{\vect{k}} c_{\vect{k}\tau}e^{i \vect{k}\cdot \vect{R}_n}$, where $N_N$ is the number of NM unit cells and $\vect{k}$ runs over the 1BZ. 
Below, we consider a spatially uniform exchange coupling, i.e., $J_{\mathrm{sd},i}=J_{\mathrm{sd}}$, and disregard Umklapp scattering between magnons and carriers. 
Furthermore, since the magnon wavelength is much larger than the lattice constant, we approximate $\exp(-i\Delta\vect{k}\cdot\vect{\iota}_i)\approx 1$, 
where $\vect{\iota}_i$ is the vector pointing from lattice site $i$ in the AFM to the center of the nearest Wannier orbital $\Psi_{\vect{R}_{i}}(\vect{r})$ in the NM and $\hbar\Delta\vect{k}$ denotes the typical momentum transfer experienced by electrons when scattering off a magnon. This approximation effectively amounts to neglecting the small spatial misalignment between the spin $\vect{S}_{i}$ and the center of the Wannier orbital $\Psi_{\vect{R}_{i}}(\vect{r})$ in the Fourier transform of Eq.~\eqref{eq:discreteinteraction}.  

The Heisenberg equation $\dot{\vect{s}}_{\rm Tot}= (i/\hbar) [\mathcal{H}, \vect{s}_{\rm Tot} ]$, with $\vect{s}_{\rm Tot}=\sum_i \vect{s}_i$, gives the rate of change of the total spin in the NM due to the coupling to the thermally excited magnons the AFM. 
The dc spin current pumped into the NM is then $\vect{I}_\mathrm{s}=  (i/\hbar) \langle [\mathcal{H}, \vect{s}_{\rm Tot} ]  \rangle $,
where the statistical average $\langle ... \rangle$  is evaluated by treating $\mathcal{H}_I$ perturbatively within linear response. 
Applying the procedure outlined in Ref.~\cite{Hals:PRB2025} and considering the weak-damping limit of the AFM, we obtain
\begin{equation}\label{eq:ultimate}
\vect{I}_\mathrm{s}= \mathcal{K}\sum_{n, \vect{q} }\vect{\Omega}_{n;\vect{q}}f_{n;\vect{q}}\left( \coth\frac{\varepsilon_n(\vect{q})}{2k_BT_\mathrm{N}} - \coth\frac{\varepsilon_n(\vect{q})}{2k_BT_\mathrm{AF}} \right) ,
\end{equation}
where $n$ runs over all magnon energy states $\varepsilon_n(\vect{q})=\hbar \omega_n (\vect{q})$ with wave vector $\vect{q}$ confined to the $xy$-plane.
The prefactor $\mathcal{K}=(L_z\chi)/(4 \hbar\lambda)$ depends on the NM-layer thickness $L_z$, the paramagnetic susceptibility $\chi$, and the spin-diffusion length $\lambda$ of the NM.
The quantities $\vect{\Omega}_{n;\vect{q}}$ and $f_{n;\vect{q}}$ are defined by
\begin{eqnarray}
\vect{\Omega}_{n;\vect{q}}&=&\sum_{\alpha,\beta}\big(  \vect{F}_{\alpha\beta;n,\vect{q}}^{+ -}  + \vect{F}_{\alpha\beta;n,\vect{q}}^{- +}  + 2 \vect{F}_{\alpha\beta;n,\vect{q}}^{z z}   \big) , \\
f_{n;\vect{q}}&=&\sqrt{\frac{\sqrt{(1+(\lambda \vect{q})^2)^2+(\omega_n(\vect{q})\tau_s )^2}-1-(\lambda \vect{q})^2}{2\left((1+(\lambda \vect{q})^2)^2+(\omega_n(\vect{q})\tau_s )^2\right)}} , \nonumber 
\end{eqnarray}
with $f_{n;\vect{q}}$ approaching zero for large $\vect{q}$.
Here, the indices $\alpha,\beta\in \{1,2,3\}$ run over the AFM sublattices. The expressions phenomenologically incorporate the spin-flip relaxation time $\tau_{s}$ and the spin-diffusion length in order to capture dissipative processes and disorder in the NM.
We have further introduced $\vect{F}_{\alpha\beta;n,\vect{q}}^{\eta \tilde{\eta} } = \vect{\Lambda}_{\alpha n;\vect{q}}^{\eta}\Gamma_{\beta n;\vect{q}}^{\tilde{\eta},*}$ ($\eta, \tilde{\eta}\in\{\pm,z\}$), where the $\Lambda$ and $\Gamma$ tensors are expressed in terms of the paraunitary matrix $\mathbf{T}_\vect{q}^{-1}$ as
\begin{eqnarray}
\vect{\Lambda}_{\alpha n;\vect{q}}^\eta &=& i\tilde{J}_\mathrm{sd}(\vect{r}_\alpha^{-\eta}U_{\alpha n,\vect{q}}+\vect{r}_\alpha^{+\eta}V_{\alpha n,\vect{q}})\text{, }  \nonumber \\
\Gamma_{\beta n;\vect{q}}^{\eta,*} &=& \frac{\tilde{J}_\mathrm{sd}}{2}(c_\beta^{-\eta}V_{\beta n,\vect{q}}^*+c_\beta^{+\eta}U_{\beta n,\vect{q}}^*) .\nonumber
\end{eqnarray}
The coupling constant is defined as $\tilde{J}_\mathrm{sd}=\hbar J_\mathrm{sd} \sqrt{SN_\mathrm{AF}/2N_\mathrm{N}} $, whereas 
the vectors are given by $\vect{r}_\alpha^{\pm\pm}=\vect{r}_\alpha^\pm\times\vect{r}^\pm$, $\vect{r}_\alpha^{\pm z}=\vect{r}_\alpha^\pm\times\hat{\vect{z}}$, 
$c_\alpha^{\pm\pm}=\vect{r}_\alpha^\pm\cdot\vect{r}^\pm$, and  $c_\alpha^{\pm z}=\vect{r}_\alpha^\pm\cdot\hat{\vect{z}}$ with 
$\vect{r}_{\alpha}^\pm=\tilde{\vect{x}}_{\alpha}^{\mathcal{C}}\pm i\tilde{\vect{y}}_{\alpha}^{\mathcal{C}}$  and $\vect{r}^\pm=\hat{\vect{x}}\pm i\hat{\vect{y}}$. Here, $\mathcal{C}=\pm$ for the 
$(\pm)$-chiral ground state. 

Eq.~\eqref{eq:ultimate} gives a microscopic expression for the thermally pumped dc spin current across the AFM/NM interface, when the AFM and NM are held at temperatures $T_\mathrm{AF}$ and $T_\mathrm{N}$, respectively. As expected, the spin current vanishes, $|\vect{I}_\mathrm{s}|=0$, when $T_\mathrm{AF}=T_\mathrm{N}$.  Analogous derivations have been performed for heterostructures of ferromagnets and collinear AFMs~\cite{Adachi:prb2011, Ohnuma:prb2013}.

Fig.~\ref{Fig3}b displays the dependence of the spin-current components on the out-of-plane magnetic field $B$ in both chiral states at $T_\mathrm{AF}=300$~K and $T_\mathrm{N}=299$~K.
The $x$-components $I_{\mathrm{s},x}^\pm$ and the $y$-component in the $(+)$-chiral state, $I_{\mathrm{s},y}^+$, vanish and are therefore omitted. 
All calculations are performed using $\tau_s=50$~fs and $\lambda=5.21$~nm, as measured in platinum at room temperature~\cite{Nair:PRL2021}.
For the AFM, we use the material parameters given in Sec.~\ref{Sec2}.

We find that the $y$-component in the $(-)$-chiral state, $I_{\mathrm{s},y}^-$, remains finite and constant for all values of the $B$-field. This behavior arises from the weak in-plane tilting by the angle $\alpha$ produced by the easy axes.
In contrast, the $z$-components, $I_{\mathrm{s},z}^\pm$, increase linearly with $B$, with $I_{\mathrm{s},z}^-$ exceeding $I_{\mathrm{s},z}^+$. 
The disparity between $I_{\mathrm{s},z}^-$ and $I_{\mathrm{s},z}^+$ stems from the (nearly) gapless excitations present in the $(-)$-chiral state.
Together, these findings establish two distinct signatures of the ground-state vector chirality: (i) a finite $y$-component of the SSE current in the $(-)$-chiral state, and (ii) an enhanced $z$-component of the spin current in the same state.
In the following, we discuss possible experimental routes to detect these signals.

\section{Experimental signatures}\label{Sec4}
The $y$-component of the pumped spin current in the $(-)$-chiral state can be detected via the inverse spin Hall effect (ISHE) in the NM.
To estimate the strength of the ISHE signal, we consider a kagome AFM monolayer interfaced with a platinum layer of thickness $L_z=10-100$~nm, width $L_y=1$~cm, and length $L_x=1$~cm, having a cubic lattice constant of $a_\mathrm{Pt}=3.9236$~Å, a Fermi energy of $\varepsilon_\mathrm{F}=0.6335R_\mathrm{H}$ \cite{Dye:JLTP1938} (where $R_\mathrm{H}$ is the Rydberg constant), and an effective charge carrier mass of $m^\ast=2.5m_\mathrm{e}$~\cite{Ketterson:PRB1970}, where $m_\mathrm{e}$ is the free electron mass.
We model the band structure of the NM by a quadratic dispersion relation for the charge carriers and use the following expression for the paramagnetic susceptibility (Appendix~\ref{app:chi}):
$\chi= (a_\mathrm{Pt}^3/\pi^2 )\sqrt{m^{\ast3}\varepsilon_\mathrm{F}/2}$.
Furthermore, using the material parameters from Sec.~\ref{Sec2} for $\text{Mn}_3\text{Ir}$,  and assuming an interface exchange interaction of $J_{\rm sd}\sim10$~meV/$\hbar^2$, 
we obtain $| I_{\mathrm{s},y}^-|= 2.7\cdot10^{18}$~$\hbar/$s for the constant $y$-component of the spin current in the ($-$)-chiral state.
This spin current generates a voltage $V_\mathrm{ISHE}=E_\mathrm{ISHE} L_x$ across the NM, where
$E_\mathrm{ISHE}=  | I_{\mathrm{s},y} | [(2e\theta_\mathrm{H}\lambda\rho)/(\hbar V) ] \tanh (L_z/2\lambda )$~\cite{Lund:PRB2021}.
Here, $\theta_\mathrm{H}$ is the spin Hall angle, $\rho$ the electric resistivity, and $V$ the volume of the NM.
For platinum at room temperature, $\lambda=5.21$~nm, $\theta_\mathrm{H}\approx0.0402$ and $\rho=10.8~\mu\Omega$cm~\cite{Nair:PRL2021}. 
From these parameters, we estimate a measurable ISHE voltage of about $V_\mathrm{ISHE} \approx 144 - 19$~nV for an NM thickness in the range of $L_z=10-100$~nm. 

Similarly, the $z$-component of the spin current can be detected by interfacing the AFM with a magnetic metal and exploiting the anomalous inverse spin Hall effect (AISHE)~\cite{Davidson:PLA2020, Wang:CP2021}. 
This effect is typically about an order of magnitude weaker than the conventional ISHE~\cite{Chuang:PLM2020, Yagmur:PLB2021, Yang:PLB2022}, which means that a ratio of $| I_{s,z}^{\pm}/I_{s,y}^- | \sim 1$ is required to obtain an AISHE voltage in the nanovolt range --- achievable under sufficiently strong magnetic fields. For the two chiral states, one can expect a relative difference in the AISHE voltages on the order of $(I_{s,z}^{-} - I_{s,z}^{+})/I_{s,z}^{+} \sim 81\%$.

The expected ISHE and IASHE voltages can be further enhanced by applying a larger temperature gradient across the AFM/NM interface (e.g., $T_\mathrm{AF}-T_\mathrm{N}=10$ K instead of 1 K), and increasing the sample length along the measurement direction.

\section{Summary and Conclusions}\label{Sec5}
In conclusion, we have demonstrated that the ground-state vector spin chirality of a kagome AFM can be detected through measurements of the SSE. In the ($-$)-chiral state, the thermally pumped spin current develops a sizable in-plane component, which is absent in the ($+$)-chiral state. This component originates from an anisotropy-induced sublattice canting and is independent of the applied out-of-plane magnetic field. We also find a field-dependent out-of-plane spin current, whose magnitude differs between the two chiralities by about 81\% due to their distinct magnon band structures. Together, the in-plane and out-of-plane SSE components provide clear signatures of the ground-state chirality, enabling real-time detection of chirality switching.

\section{Acknowledgements}
We thank Mathias Kläui for stimulating discussions.
KMDH acknowledges funding from the Research Council of Norway via Project No. 334202.


\appendix
\section{The in-plane tilting angle $\alpha$}\label{app:alpha}
To minimize the energy of the spin Hamiltonian $\mathcal{H}_\text{AF}^-$ of the AFM in the $(-)$-chiral state [Eq.~\eqref{eq:ham}] with respect to the in-plane tilting angle $\alpha$, we insert the ansatz
$\vect{S}_i=\hbar S\tilde{\vect{n}}_i^-$, with $\tilde{\vect{n}}_1^-=(0,-1,0)$, $\tilde{\vect{n}}_2^-=(\cos(\pi/6+\alpha),\sin(\pi/6+\alpha),0)$, and $\tilde{\vect{n}}_3^-=(-\cos(\pi/6+\alpha),\sin(\pi/6+\alpha),0)$ for the spins, into Eq.~\eqref{eq:ham}. For $B=0$, $\mathcal{H}_{\text{AF}}^-$ then reduces to
\begin{equation}
\begin{split}
\mathcal{H}_\text{AF}^- =&\bigg(2J\big(-\cos\alpha-\sqrt{3}\sin\alpha+\sqrt{3}\sin\alpha\cos\alpha\\
&\qquad+\frac{1}{2}(\sin^2\alpha-\cos^2\alpha)\big)\\
&-2D\big(\sqrt{3}\cos\alpha-\sin\alpha+\sin\alpha\cos\alpha\\
&\qquad+\frac{\sqrt{3}}{2}(\cos^2\alpha-\sin^2\alpha)\big)\\
&-K\big(1+\frac{\cos^2\alpha}{2}-\sqrt{3}\sin\alpha\cos\alpha+\frac{3}{2}\sin\alpha\big)\bigg)\\
&\times N_\mathrm{AF} S^2\hbar^2 \text{.}
\end{split}
 \label{eq:unit-}
\end{equation}
Differentiating with respect to $\alpha$ and expanding for small angles, i.e., $\sin\alpha\approx\alpha$ and $\cos\alpha\approx1$, we obtain
\begin{equation}
\alpha \approx -\frac{\sqrt{3}K}{2(3J+3\sqrt{3}D-K)} < 0\text{,}
\label{eq:alpha}
\end{equation}
which remains finite as long as the in-plane easy-axis anisotropy is nonzero.

\section{The out-of-plane tilting angle $\theta$}\label{app:theta}
In order to minimize the spin Hamiltonian $\mathcal{H}_\text{AF}$ in Eq.~\eqref{eq:ham} with respect to the out-of-plane tilting angle $\theta$, we employ the ansatz
$\vect{S}_i= \hbar S[ \cos \theta \tilde{\vect{n}}_i + \sin \theta \hat{\vect{z}}]$. Substituting $\vect{S}_i$ into Eq.~\eqref{eq:ham}, $\mathcal{H}_\text{AF}$ becomes
\begin{equation}
\begin{split}
\mathcal{H}_\text{AF}^+ =&3\big(\hbar SJ(3\sin^2\theta-1)+\hbar S(K_\perp\sin^2\theta-K\cos^2\theta)\\
&+\sqrt{3}\hbar SD\cos^2\theta-3\gamma B\sin\theta\big)N_\mathrm{AF}\hbar S\text{.}
\end{split}
 \label{eq:unit}
\end{equation}
Differentiation with respect to the tilting angle yields the expression for $\theta$ in the $(+)$-chiral ground state:
\begin{equation}
\theta^+=\arcsin\frac{\gamma B}{2\hbar S(3J+K_\perp+K-\sqrt{3}D)}\text{.}
 \label{eq:theta+}
\end{equation}
As expected, $\theta$ vanishes in the limit $B\rightarrow 0$. Similarly, the tilting angle for the $(-)$-chiral configuration becomes
\begin{equation}
\theta^-=\arcsin\frac{\gamma B}{2\hbar S(3J+K_\perp+\frac{K}{2}+\sqrt{3}D)}\text{,}
 \label{eq:theta-}
\end{equation}
with $\theta^+=\theta^-$ at the critical value $D=\frac{K}{4\sqrt{3}}$.

\section{The BdG Hamiltonian}\label{app:bdg}
The $6\times 6$ Bogoliubov-de-Gennes (BdG) matrix $\mathbf{H}_{\vect{q}}$ from Eq.~\eqref{eq:bdg} in the main text has the form
\begin{equation}\label{eq:matrix}
    \mathbf{H}_{\vect{q}}^+=
    \begin{pmatrix}
    \Gamma & \Lambda_1& \Lambda_3^* & \tilde{\Gamma} & \tilde{\Lambda}_1 & \tilde{\Lambda}_3\\
    \Lambda_1^* & \Gamma & \Lambda_2 & \tilde{\Lambda}_1 & \tilde{\Gamma} & \tilde{\Lambda}_2\\
    \Lambda_3 & \Lambda_2^* & \Gamma & \tilde{\Lambda}_3 & \tilde{\Lambda}_2& \tilde{\Gamma}\\
    \tilde{\Gamma} & \tilde{\Lambda}_1 & \tilde{\Lambda}_3 & \Gamma & \Lambda_1^* & \Lambda_3\\
    \tilde{\Lambda}_1 & \tilde{\Gamma} & \tilde{\Lambda}_2 & \Lambda_1 & \Gamma & \Lambda_2^*\\
    \tilde{\Lambda}_3 & \tilde{\Lambda}_2 & \tilde{\Gamma} & \Lambda_3^*& \Lambda_2 & \Gamma
    \end{pmatrix}
    \end{equation}
for the $(+)$-chiral configuration, where
\begin{eqnarray}\label{eq:whole}
&&\overset{(\sim)}{\Lambda}_i=\overset{(\sim)}{\Lambda}\cos(\vect{q}\cdot a\hat{\vect{e}}_i),\quad i\in\{1,2,3\}\nonumber\\
&&\Gamma=  S\hbar^2\big((2J+K_\perp)(1-3\sin^2\theta^+)+K(2-3\sin^2\theta^+)\nonumber\\
&&\qquad\qquad-2\sqrt{3}D\cos^2\theta^+\big)+\hbar\gamma B\sin\theta^+, \nonumber\\
&&\tilde{\Gamma}=-S\hbar^2(K_\perp\cos^2\theta^+-K\sin^2\theta^+)\text{,}\nonumber\\
&&\Lambda =S\hbar^2\Bigg(-J\left(\sin^2\theta^+-\frac{\cos^2\theta^+}{2}+i\sqrt{3}\sin\theta^+\right)\\
&&\qquad\qquad+D\left(\frac{\sqrt{3}}{2}+\frac{\sqrt{3}}{2}\sin^2\theta^+-i\sin\theta^+\right)\Bigg)\text{,}\nonumber\\
&&\tilde{\Lambda} =-\frac{\sqrt{3}}{2}S\hbar^{2}(\sqrt{3}J-D)\cos^2\theta^+\text{,}\nonumber
\end{eqnarray}
and the asterisk (*) denotes complex conjugation. The BdG Hamiltonian for the $(-)$-chiral configuration is

\begin{equation}\label{eq:matrix-}
    \mathbf{H}_{\vect{q}}^-=
    \begin{pmatrix}
    \bar{\Gamma}_1 & \bar{\Lambda}_{12} & \bar{\Lambda}_{31}^* & \tilde{\Gamma} & \hat{\Lambda}_{12} & \hat{\Lambda}_{31}\\
    \bar{\Lambda}_{12}^* & \bar{\Gamma}_2 & \bar{\Lambda}_{23} & \hat{\Lambda}_{12} & \hat{\Gamma}^* & \hat{\Lambda}_{23}\\
    \bar{\Lambda}_{31} & \bar{\Lambda}_{23}^* & \bar{\Gamma}_3 & \hat{\Lambda}_{31} & \hat{\Lambda}_{23}& \hat{\Gamma}\\
    \tilde{\Gamma} & \hat{\Lambda}_{12} & \hat{\Lambda}_{31} & \bar{\Gamma}_1 & \bar{\Lambda}_{12}^* & \bar{\Lambda}_{31}\\
    \hat{\Lambda}_{12} & \hat{\Gamma} & \hat{\Lambda}_{23} & \bar{\Lambda}_{12} & \bar{\Gamma}_2& \bar{\Lambda}_{23}^*\\
    \hat{\Lambda}_{31} & \hat{\Lambda}_{23} & \hat{\Gamma}^* & \bar{\Lambda}_{31}^*& \bar{\Lambda}_{23} & \bar{\Gamma}_3
    \end{pmatrix}
    \text{,}
    \end{equation}
where the matrix entries are defined as
\begin{eqnarray}\label{eq:wholenew}
&&\bar{\Lambda}_{ij}=\Pi_{ij}\cos(\vect{q}\cdot a\hat{\vect{e}}_i),\quad i,j\in\{1,2,3\}\nonumber\\
&&\hat{\Lambda}_{ij}=Y_{ij}\cos(\vect{q}\cdot a\hat{\vect{e}}_i),\quad i,j\in\{1,2,3\}\nonumber\\
&&\bar{\Gamma}_i=M_i+S\hbar^2K_\perp(1-3\sin^2\theta^-)+\hbar\gamma B\sin\theta^-,\nonumber\\
&&M_1=-2S\hbar^2\big(J\left((n_{12}+n_{31})\cos^2\theta^-+2\sin^2\theta^-\right)\nonumber\\
&&\qquad\qquad\qquad+D(z_{12}+z_{31})\cos^2\theta^-\big)+C, \nonumber\\
&&M_2=-2S\hbar^2\big(J\left((n_{12}+n_{23})\cos^2\theta^-+2\sin^2\theta^-\right)\nonumber\\
&&\qquad\qquad\qquad+D(z_{12}+z_{23})\cos^2\theta^-\big)+\bar{C}, \nonumber\\
&&M_3=-2S\hbar^2\big(J\left((n_{31}+n_{31})\cos^2\theta^-+2\sin^2\theta^-\right)\\
&&\qquad\qquad\qquad+D(z_{31}+z_{31})\cos^2\theta^-\big)+\bar{C}, \nonumber\\
&&C=S\hbar^2K(2-3\sin^2\theta^-),\nonumber\\
&&\bar{C}=S\hbar^2K\left(\cos^2\left(\frac{\pi}{3}+\alpha\right)(2-3\sin^2\theta^-)-\sin^2\left(\frac{\pi}{3}+\alpha\right)\right),\nonumber\\
&&\tilde{\Gamma}=-S\hbar^2(K_\perp\cos^2\theta^--K\sin^2\theta^-)\text{,}\nonumber\\
&&\hat{\Gamma}=S\hbar^2\bigg(-K_\perp\cos^2\theta^-+K\bigg(\cos^2\left(\frac{\pi}{3}+\alpha\right)\sin^2\theta^-\nonumber\\
&&\qquad\qquad-\sin^2\left(\frac{\pi}{3}+\alpha\right)+i\sin\left(\frac{2\pi}{3}+2\alpha\right)\sin\theta^-\bigg)\bigg)\text{,}\nonumber\\
&&\Pi_{ij} =S\hbar^2\bigg(J\left(n_{ij}(1+\sin^2\theta^-)+\cos^2\theta^-+2iz_{ij}\sin\theta^-\right)\nonumber\\
&&\qquad\qquad+D\left(z_{ij}(1+\sin^2\theta^-)-2in_{ij}\sin\theta^-\right)\bigg)\text{,}\nonumber\\
&&Y_{ij}=S\hbar^{2}\left(J(n_{ij}-1)+Dz_{ij}\right)\cos^2\theta^-\text{,}\nonumber\\
&&n_{ij}=\tilde{\vect{n}}_i^-\cdot\tilde{\vect{n}}_j^-\text{,}\quad z_{ij}=-\hat{\vect{z}}\cdot(\tilde{\vect{n}}_i^-\times\tilde{\vect{n}}_j^-)\text{.}\nonumber
\end{eqnarray}

\section{Spin susceptibility of NM}\label{app:chi}
The spin susceptibility is defined through the retarded Green’s function  $\chi (\vect{r}-\vect{r}^{'}, t-t^{'})= -i \theta (t-t^{'}) \langle [s^{-} (\vect{r}, t), s^{+} (\vect{r}^{'}, t^{'})] \rangle$.
Here, $s^{\pm}= s_x\pm s_y$ where $\vect{s}(\vect{r},t)$ is the spin density of the
carriers in the Heisenberg picture. In frequency and momentum space, the Green's function takes the form~\cite{Doniach:book}
\begin{equation}
\chi (\vect{k}, \omega) =  \frac{\hbar^2}{N_N} \sum_{\vect{p}} \frac{f(\epsilon_{\vect{p}}) -  f(\epsilon_{\vect{p} +\vect{k}}) }{\omega -  (\epsilon_{\vect{p}} - \epsilon_{\vect{p}+ \vect{k}})/\hbar + i 0^{+}} .
\end{equation}
For $\vect{k}\to 0$ and $\omega \to 0$, we obtain $\chi=\hbar^3\mathcal{N}(\varepsilon_\mathrm{F})$, where
\begin{equation}
\mathcal{N}(\varepsilon)=\frac{1}{N_\mathrm{N}}\sum_{\vect{k}}\delta(\varepsilon-\varepsilon_{\vect{k}})
\end{equation}
is the single-particle density of states. Rewriting the sum as an integral, $\sum_{\vect{k}}\rightarrow V/(2\pi)^3\int d^3k$ , and assuming a quadratic dispersion $\varepsilon_{\vect{k}}=\hbar^2 k^2/2m^\ast$, we obtain
\begin{equation}
\mathcal{N}(\varepsilon_\mathrm{F})=\frac{V_\mathrm{u.c.}}{\pi^2\hbar^3}\sqrt{\frac{m^{\ast3}\varepsilon_\mathrm{F}}{2}},
\end{equation}
where $V_\mathrm{u.c.}$ is the NM unit cell volume. Substituting this into the susceptibility and using $V_\mathrm{u.c.}=a_\mathrm{Pt}^3$ for platinum yields the result given in Sec.~\ref{Sec4}.



\begin{thebibliography}{99} 

\bibitem{Sitte:jap2018}K. Everschor-Sitte, J. Masell, R. M. Reeve, and  M. Kl\"aui, Perspective: Magnetic skyrmions -- Overview of recent progress in an active research field,  J. Appl. Phys. {\bf 124}, 240901 (2018).
\bibitem{Bobel:pr2021}B. G\"obel, I. Mertig, and O. A.Tretiakov, Beyond skyrmions: Review and perspectives of alternative magnetic quasiparticles, Physics Reports {\bf 895}, 1–28 (2021).
\bibitem{Manchon:ssp2017}A. Manchon and A. Belabbes, Spin-Orbitronics at Transition Metal Interfaces, Solid State Physics {\bf 68}, 1-89 (2017).
\bibitem{Brataas:nn2014}A. Brataas and K. M. D. Hals, Spin–orbit torques in action, Nature Nanotechnology {\bf 9}, 86 (2014).

\bibitem{Andreev:spu1980} A. F. Andreev and V. I. Marchenko, Symmetry and the macroscopic dynamics of magnetic materials, Sov. Phys. Usp. {\bf 23}, 21 (1980).

\bibitem{Smejkal:nrm2022}L. Šmejkal, A. H. MacDonald, J. Sinova, S. Nakatsuji, and T. Jungwirth, Anomalous Hall antiferromagnets,  Nature Reviews Materials {\bf 7}, 482 (2022).

\bibitem{Nakatsuji:nature2015}S. Nakatsuji,  N. Kiyohara,  and T. Higo,  Large anomalous Hall effect in a non-collinear antiferromagnet at room temperature, Nature {\bf 527}, 212 (2015).
\bibitem{Nayak:Sadv2016}A. K. Nayak, J. E. Fischer, Y. Sun, B. Yan, J. Karel, A. C. Komarek, C. Shekhar, N. Kumar, W. Schnelle, J. Kübler, C. Felser, and S. S. P. Parkin,  Large anomalous Hall effect driven by a nonvanishing Berry curvature in the noncolinear antiferromagnet Mn$_3$Ge, Sci. Adv. {\bf 2}, e1501870 (2016).
\bibitem{Kiyohara:pra2015} N. Kiyohara,  T. Tomita,  and S. Nakatsuji, Giant anomalous Hall effect in the chiral antiferromagnet $\text{Mn}_3\text{Ge}$, Phys. Rev. Appl. {\bf 5}, 064009 (2016).

\bibitem{Ikhlas:nphys2017}M. Ikhlas, T. Tomita, T. Koretsune, M.-T. Suzuki, D. Nishio-Hamane, R. Arita, Y. Otani, S. Nakatsuji, Large anomalous Nernst effect at room temperature in a chiral antiferromagnet, Nat. Phys. {\bf 13}, 1085 (2017).

\bibitem{Higo:np2018}T. Higo, H. Man, D. B. Gopman, L. Wu, T. Koretsune, O. M. J. van ’t Erve, Y. P. Kabanov, D. Rees, Y. Li, M.-T. Suzuki, S. Patankar, M. Ikhlas, C. L. Chien, R. Arita, R. D. Shull, J. Orenstein, and Satoru Nakatsuji, Large magneto-optical Kerr effect and imaging of magnetic octupole domains in an antiferromagnetic metal, Nature Photonics {\bf 12}, 73 (2018).

\bibitem{Kimata:n2019}M. Kimata, H. Chen, K. Kondou, S. Sugimoto, P. K. Muduli, M. Ikhlas, Y. Omori, T. Tomita, A. H. MacDonald, S. Nakatsuji, and Y. Otani, Magnetic and magnetic inverse spin Hall effects in a non-collinear antiferromagnet, Nature {\bf 565}, 627 (2019).
\bibitem{Sallermann:prl2022}D. Go, M. Sallermann, F. R. Lux, S. Bl\"ugel, O. Gomonay, and Y. Mokrousov, Noncollinear Spin Current for Switching of Chiral Magnetic Textures, Phys. Rev. Lett. {\bf 129}, 097204 (2022).

\bibitem{Grohol:nm2005}D. Grohol, K. Matan, J.-H. Cho, S.-H. Lee, J. W. Lynn, D. G. Nocera and Y. S. Lee, Spin chirality on a two-dimensional frustrated lattice, Nat. Mat.  {\bf 4},  323 (2005).

\bibitem{Kuroda:natmat2017}K. Kuroda, T. Tomita, M.-T. Suzuki, C. Bareille, A. A. Nugroho, P. Goswami, M. Ochi, M. Ikhlas, M. Nakayama, S. Akebi, R. Noguchi, R. Ishii, N. Inami, K. Ono, H. Kumigashira, A. Varykhalov, T. Muro, T. Koretsune, R. Arita, S. Shin, Takeshi Kondo, and S. Nakatsuji,  Evidence for magnetic Weyl fermions in a correlated metal, Nat. Mater. {\bf 16}, 1090 (2017).

\bibitem{Jungwirth:np2018} T. Jungwirth, J. Sinova, A. Manchon, X. Marti, J. Wunderlich and C. Felser, The multiple directions of antiferromagnetic spintronics, Nat. Phys.  {\bf 14}, 200 (2018).
\bibitem{Duine:np2018} R. A. Duine, Kyung-Jin Lee, S. P. Parkin and M. D. Stiles, Synthetic antiferromagnetic spintronics, Nat. Phys. {\bf 14}, 217 (2018).
\bibitem{Gomonay:np2018} O. Gomonay, V. Baltz, A. Brataas and Y. Tserkovnyak, Antiferromagnetic spin textures and dynamics, Nat. Phys. {\bf 14}, 213 (2018). 
\bibitem{Zelezny:np2018} J. {\v Z}elezn{\'y}, P. Wadley, K. Olejn{\'i}k, A. Hoffmann and H. Ohno, Spin transport and spin torque in antiferromagnetic devices, Nat. Phys. {\bf 14}, 220 (2018).
\bibitem{Nemec:np2018} P. N{\v e}mec, M. Fiebig, T. Kampfrath and A. V. Kimel, Antiferromagnetic opto-spintronics, Nat. Phys. {\bf 14}, 229 (2018).
\bibitem{Libor:np2018} L. {\v S}mejkal, Y. Mokrousov, B. Yan and A. H. MacDonald, Topological antiferromagnetic spintronics, Nat. Phys. {\bf 14}, 242 (2018).
\bibitem{Baltz:rmp2018}V. Baltz, A. Manchon, M. Tsoi, T. Moriyama, T. Ono, and Y. Tserkovnyak, Antiferromagnetic spintronics, Rev. Mod. Phys. {\bf 90}, 015005 (2018).

\bibitem{Gomonay:PRB2012}H. V. Gomonay, R. V. Kunitsyn, and V. M. Loktev, Symmetry and the macroscopic dynamics of antiferromagnetic materials in the presence of spin-polarized current, Phys. Rev. B {\bf 85}, 134446 (2012).
\bibitem{Tserkovnyak:PRB2017}Y. Tserkovnyak and H. Ochoa, Generalized boundary conditions for spin transfer, Phys. Rev. B {\bf 96}, 100402 (2017).
\bibitem{Ochoa:PRB2018}H. Ochoa, R. Zarzuela, and Y. Tserkovnyak, Spin hydrodynamics in amorphous magnets, Phys. Rev. B {\bf 98}, 054424 (2018).
\bibitem{Yamane:PRB2019}Y. Yamane, O. Gomonay, and J. Sinova, Dynamics of noncollinear antiferromagnetic textures driven by spin current injection, Phys. Rev. B {\bf 100}, 054415 (2019).
\bibitem{Tsai:nature2020}H. Tsai, T. Higo, K. Kondou, T. Nomoto, A. Sakai, A. Kobayashi, T. Nakano, K. Yakushiji, R. Arita, S. Miwa, Y. Otani, and S. Nakatsuji , Electrical manipulation of a topological antiferromagnetic state, Nature {\bf 580}, 608 (2020).
\bibitem{Takeuchi:nmat2021}Y. Takeuchi, Y. Yamane, J.Y.Yoon, R. Itoh, B. Jinnai, S. Kanai, J. Ieda, S. Fukami, and H. Ohno, Chiral-spin rotation of non-collinear antiferromagnet by spin–orbit torque, Nature Materials, {\bf 20},  1364 (2021). 
\bibitem{Li:PRB2021}B. Li and A. A. Kovalev, Spin superfluidity in noncollinear antiferromagnets, Phys. Rev. B {\bf 103}, L060406 (2021).
\bibitem{Lund:PRB2021}M. A. Lund, A. Salimath, and K. M. D. Hals, Spin pumping in noncollinear antiferromagnets, Phys. Rev. B {\bf 104}, 174424 (2021).

\bibitem{Szunyogh:PRB2009} L. Szunyogh, B. Lazarovits, L. Udvardi, J. Jackson, and U. Nowak, Giant magnetic anisotropy of the bulk antiferromagnets IrMn and $\text{IrMn}_3$ from first principles, Phys. Rev. B {\bf 79}, 020403(R) (2009).
\bibitem{Jenkins:JAP2018}Sarah Jenkins and Richard F. L. Evans, Enhanced finite size and interface mixing effects in iridium manganese ultra thin films, Journal of Applied Physics 124, 152105 (2018).

\bibitem{Xie:nc2022}H. Xie, X. Chen, Q. Zhang, Z. Mu, X. Zhang, B. Yan, and Y. Wu, Magnetization switching in polycrystalline $\text{Mn}_3\text{Sn}$ thin film induced by self-generated spin-polarized current, Nat Commun {\bf 13}, 5744 (2022).
\bibitem{Yan:am2022}G. Q. Yan, S. Li, H. Lu, M. Huang, Y. Xiao, L. Wernert, J. A. Brock, E. E. Fullerton, H. Chen, H. Wang, and C. R. Du, Quantum Sensing and Imaging of Spin–Orbit-Torque-Driven Spin Dynamics in the Non-Collinear Antiferromagnet $\text{Mn}_3\text{Sn}$, Advanced Materials {\bf 34}, 2200327 (2022).
\bibitem{Higo:Nature2022}T. Higo, K. Kondou, T. Nomoto, M. Shiga, S. Sakamoto, X. Chen, D. Nishio-Hamane, R. Arita, Y. Otani, S. Miwa, and S. Nakatsuji, Perpendicular full switching of chiral antiferromagnetic order by current, Nature {\bf 607}, 474 (2022).
\bibitem{Rodrigues:PRB2022}D. R. Rodrigues, A. Salimath, K. Everschor-Sitte, and K. M. D. Hals, Dzyaloshinskii-Moriya induced spin-transfer torques in kagome antiferromagnets, Phys. Rev. B {\bf 105}, 174401 (2022).
\bibitem{Yoon:nmat2023}J.-Y. Yoon, P. Zhang, C.-T. Chou, Y. Takeuchi, T. Uchimura, J. T. Hou, J. Han, S. Kanai, H. Ohno, S. Fukami, and L. Liu, Handedness anomaly in a non-collinear antiferromagnet under spin–orbit torque, Nat. Mater. {\bf 22}, 1106 (2023).
\bibitem{Lund:PRB2023}M. A. Lund, D. R. Rodrigues, K. Everschor-Sitte, and K. M. D. Hals, Voltage-Controlled High-Bandwidth Terahertz Oscillators Based on Antiferromagnets, Phys. Rev. Lett. {\bf 131}, 156704 (2023).
\bibitem{Comment}In this limit, the lowest magnon band of the $(-)$-chiral state becomes gapless at the $\Gamma$-point.
\bibitem{Deng:nsr2023}Y. Deng, X. Liu, Y. Chen, Z. Du, N. Jiang, C. Shen, E. Zhang, H. Zheng,H. Z. Lu , and K. Wang, All-electrical switching of a topological non-collinear antiferromagnet at room temperature, Natl. Sci Rev, {\bf10}, nwac154 (2023). 
\bibitem{Lee:nn2025}W.-B. Lee, S. Hwang, H.-W. Ko, B.-G. Park, K.-J. Lee, and G.-M. Choi, Spin-torque-driven gigahertz magnetization dynamics in the non-collinear antiferromagnet $\text{Mn}_3\text{Sn}$, Nat. Nanotechnol. {\bf 20}, 487 (2025).

\bibitem{Tharmalingam:prb2025}M. Tharmalingam, F. S. Konomaev, and K. M. D. Hals, Strain-induced manipulation of noncollinear antiferromagnets, Phys. Rev. B {\bf 112}, 104410 (2025).

\bibitem{Reichlova:nc2019}H. Reichlova, T. Janda, J. Godinho, A. Markou, D. Kriegner, R. Schlitz, J. Zelezny, Z. Soban, M. Bejarano, H. Schultheiss, P. Nemec, T. Jungwirth, C. Felser, J. Wunderlich, and S. T. B. Goennenwein, Imaging and writing magnetic domains in the non-collinear antiferromagnet Mn$_3$Sn, Nature Communications {\bf 10},  5459 (2019)

\bibitem{Rodrigues:PRL2021}D. R. Rodrigues, A. Salimath, K. Everschor-Sitte, and K. M. D. Hals, Spin-wave driven bidirectional domain wall motion in kagome antiferromagnets, Phys. Rev. Lett. {\bf 127}, 157203 (2021).

\bibitem{Chakraborty:nc2025}A. Chakraborty, A. Birk Hellenes, R. Jaeschke-Ubiergo, T. Jungwirth, L. Šmejkal, and J. Sinova, Highly efficient non-relativistic Edelstein effect in nodal p-wave magnets, Nature Communications {\bf 16}, 7270 (2025).

\bibitem{Laurell:PRB2018}P. Laurell and G.A. Fiete, Magnon thermal Hall effect in kagome antiferromagnets with Dzyaloshinskii-Moriya interactions, Phys. Rev. B {\bf 98}, 094419 (2018).
\bibitem{Hals:PRB2025}F. S. Konomaev and K. M. D. Hals, Amplifying the antiferromagnetic spin Seebeck effect through topological magnons, Phys. Rev. B {\bf 111}, 024403 (2025). 

\bibitem{Bauer:review}G. E. W. Bauer, E. Saitoh, and B. J. van Wees, Spin caloritronics, Nature Materials {\bf 11}, 391 (2012). 

\bibitem{Auerbach:book}A. Auerbach, \emph{Interacting Electrons and Quantum Magnetism} (Springer-Verlag, New York, 1994)

\bibitem{Colpa}J.H.P. Colpa, Diagonalization of the quadratic boson hamiltonian, Physica {\bf 93A},  327 (1978).

\bibitem{Tomita:JPSJ2020}T. Tomita, M. Ikhlas, and S. Nakatsuji, Large Nernst Effect and Thermodynamics Properties in Weyl Antiferromagnet, in Proceedings of the International Conference on Strongly Correlated Electron Systems (SCES2019), Vol. {\bf 30} (Journal of the Physical Society of Japan, 2020).

\bibitem{Adachi:prb2011}H. Adachi, J.-i. Ohe, S. Takahashi, and S. Maekawa, Linear-response theory of spin Seebeck effect in ferromagnetic insulators, Phys. Rev. B {\bf 83}, 094410 (2011).
\bibitem{Ohnuma:prb2013}Y. Ohnuma, H. Adachi, E. Saitoh, and S. Maekawa, Spin Seebeck effect in antiferromagnets and compensated ferrimagnets, Phys. Rev. B {\bf 87}, 014423 (2013).

\bibitem{Nair:PRL2021} R. S. Nair, E. Barati, K. Gupta, Zh. Yuan and P. J. Kelly, Spin-Flip Diffusion Length in 5d Transition Metal Elements: A First-Principles Benchmark, Phys. Rev. Lett. {\bf 126}, 196601 (2021).

\bibitem{Dye:JLTP1938}D. H. Dye, J. B. Ketterson, and G. W. Crabtree, The Fermi Surface of Platinum, Journal of Low Temperature Physics, {\bf 30}, Nos. 5/6 (1978).
\bibitem{Ketterson:PRB1970} J. B. Ketterson and L. R. Windmiller, de Haas-van Alphen Effect in Platinum, Phys. Rev. B {\bf 2}, 4813 (1970).

\bibitem{Davidson:PLA2020}A. Davidson, V. P. Amin, W. S. Aljuaid, P. M. Haney, X. Fan, Perspectives of electrically generated spin currents in ferromagnetic materials, Physics Letters A {\bf 384}, 126228 (2020).
\bibitem{Wang:CP2021}X. R. Wang, Anomalous spin Hall and inverse spin Hall effects in magnetic systems, Communications Physics {\bf 4}, 55 (2021). 

\bibitem{Chuang:PLM2020}T. C. Chuang, D. Qu, S. Y. Huang, and S. F. Lee, Magnetization-dependent spin Hall effect in a perpendicular magnetized film, Phys. Rev. Research {\bf 2}, 032053(R) (2020).
\bibitem{Yagmur:PLB2021}A. Yagmur, S. Sumi, H. Awano, and K. Tanabe, Magnetization-dependent inverse spin Hall effect in compensated ferrimagnet TbCo alloys, Phys. Rev. B {\bf 103}, 214408 (2021).
\bibitem{Yang:PLB2022}M. Yang, B. Miao, J. Cheng, K. He, X. Yang, Y. Zeng, Z. Wang, L. Sun, X. Wang, A. Azevedo, S. Bedanta, and H. Ding, Anomalous inverse spin Hall effect in perpendicularly magnetized Co/Pd multilayers, Phys. Rev. B {\bf 105}, 224426 (2022).

\bibitem{Doniach:book}S. Doniach and E. H. Sondheimer, \emph{Green's function for solid state physicists} (Imperial College Press, London, 1998).

\bibitem{data} F. S. Konomaev, M. Tharmalingam, and K. M. D. Hals, Replication Data for: Spin-Seebeck Signatures of Chirality in Kagome Antiferromagnets, DataverseNO, V2 (2025), https://doi.org/10.18710/CN2VGX.

\end{thebibliography}
\end{document}